\documentclass[12pt]{article}

\usepackage{times}
\usepackage{epsfig}
\usepackage{graphicx}
\usepackage{amsmath,amssymb}

\newcommand{\beq}[1]{
\begin{equation}
\label{e#1} }

\newcommand{\eeq}{
\end{equation}
}

\newenvironment{sciabstract}{%
\begin{quote} \bf}
{\end{quote}}

\newcounter{lastnote}

\title{Spin Hall effect transistor}

\author
{J\"{o}rg~Wunderlich,$^{1,2\ast\#}$ Byong-Guk~Park,$^{1\ast}$ Andrew~C.~Irvine$^{3\ast}$ Liviu~P.~Z\^{a}rbo,$^{2}$ \\ Eva~Rozkotov\'a,$^{4}$ Petr~N\v{e}mec,$^{4}$ V\'{\i}t Nov\'ak,$^{2}$ Jairo~Sinova,$^{5,2}$ Tom\'a\v{s}~Jungwirth$^{2,6}$\\
\\
\normalsize{$^{1}$Hitachi Cambridge Laboratory, Cambridge CB3 0HE, United Kingdom}\\
\normalsize{$^{2}$Institute of Physics ASCR, v.v.i., Cukrovarnick\'a 10, 162 53
Praha 6, Czech Republic}\\
\normalsize{$^{3}$Microelectronics Research Centre, Cavendish Laboratory,
}\\
\normalsize{
University of Cambridge, CB3 0HE, United Kingdom}\\
\normalsize{$^{4}$Faculty of Mathematics and Physics, Charles University in Prague,}\\
\normalsize{Ke Karlovu 3,
121 16 Prague 2, Czech Republic}\\
\normalsize{$^{5}$Department of Physics, Texas A\&M University, College
Station, TX 77843-4242, USA}\\
\normalsize{$^{6}$School of Physics and
Astronomy, University of Nottingham,}\\
\normalsize{Nottingham NG7 2RD, United Kingdom}\\
\\
\normalsize{$^\ast$These authors contributed equally to this work.}
\\
\normalsize{$^\#$To whom correspondence should be addressed; E-mail:  jw526@cam.ac.uk.}
}

\date{}

\begin{document}

\baselineskip24pt

\maketitle

\begin{sciabstract}

Spin transistors and spin Hall effects have been two separate leading directions of  research in semiconductor spintronics which seeks new paradigms for information processing technologies. We have brought the two directions together to realize an all-semiconductor spin Hall effect transistor.  Our scheme circumvents semiconductor-ferromagnet interface problems of the original Datta-Das  spin transistor concept and demonstrates the utility of the spin Hall effects in microelectronics. The devices use diffusive transport and operate without electrical current, i.e., without Joule heating in the active part of the transistor. We demonstrate a spin AND logic function in a  semiconductor channel with two gates. Our experimental study is complemented by numerical Monte Carlo simulations of spin-diffusion through the transistor channel. 
\end{sciabstract}

To date the two major themes in semiconductor spintronics research, the spin transistors and the spin Hall effects, have followed distinct and independent scientific paths \cite{Zutic:2004_a,Sinova:2008_a}. In the transistor case, the target device concept was established from the outset by Datta and Das \cite{Datta:1990_a}. The ensuing research has been focused on  the fundamental physical problems related  to the resistance mismatch between the semiconducting channel and the ferromagnetic spin injector and detector and to spin-manipulation  in the semiconductor via spin-orbit coupling effects \cite{Kikkawa:1999_a,Zhu:2001_a,Hammar:2002_a,Schmidt:2002_a,Schliemann:2003_c,Bernevig:2006_a,Jiang:2005_b,Crooker:2005_a,Weber:2006_a,Lou:2007_a,Huang:2007_a,Koo:2009_a}. Quite opposite for the spin Hall effect case, much of the related intriguing quantum-relativistic physics \cite{Dyakonov:1971_b,Hirsch:1999_a,Murakami:2003_a,Sinova:2004_a} has been established before the first experimental observations \cite{Kato:2004_d,Wunderlich:2004_a} but the field is still seeking basic paradigms for turning the phenomenon into a concrete device functionality. Here we demonstrate the applicability of the spin Hall effect in a new type of spin transistor.  In the device, gate-controlled output electrical signals are realized in the form of spin-dependent Hall voltages detected along a non-magnetic semiconductor channel. Our spin-Hall effect transistor shares with the Datta-Das transistor \cite{Koo:2009_a} the spin-orbit coupling based method of spin manipulation. Unlike the original Datta-Das concept, it utilizes spin-orbit coupling effects also for spin detection, completely eliminates magnetic elements from the detector, and operates with zero charge current in the detection and gated parts of the transistor channel.

Our devices comprise the active semiconductor channel in the form of a two-dimensional electron gas (2DEG) in which the spin-orbit coupling induced spin precession is controlled by external gate electrodes  and detection is provided by transverse spin Hall effect voltages \cite{Wunderlich:2008_a} measured along the 2DEG Hall bar. This spin manipulation and detection scheme  is in principle compatible with optical spin injection as well as with electrical spin-injection from a ferromagnet. In the devices presented in this paper we utilize the optical spin injection method \cite{Wunderlich:2008_a}. This way all three components of the spin transistor, i.e. spin injection, manipulation, and detection, are realized within one all-semiconductor structure and do not require any magnetic elements or external magnetic fields for the operation of the device. Because of the nondestructive nature of the spin Hall effect detection, one semiconductor channel can accommodate multiple gates and Hall cross detectors and is therefore directly suitable for realizing spin logic operations.

The semiconductor epilayers we use, described in detail in Ref.~\cite{Wunderlich:2008_a}, comprise a modulation p-doped AlGaAs/GaAs heterojunction on  top of the structure, 90~nm of intrinsic GaAs, and an n-doped AlGaAs/GaAs heterojunction underneath. In the unetched part of the wafer the top heterojunction is populated by holes while the 2DEG at the bottom heterojunction is partly or completely depleted, depending on the specific doping profile in the heterostructure. The n-side of the co-planar p-n junction is formed by removing the p-doped surface layer from a part of the wafer which results in populating  the 2DEG. At zero or reverse bias, the device is not conductive in dark due to charge depletion at the lateral p-n junction. Counter-propagating electron and hole currents can be generated by illumination at sub-gap wavelengths \cite{Wunderlich:2008_a}. Due to optical selection rules, the out-of-plane spin-polarization of injected electrons  is determined by the sense and degree of the circular polarization of vertically incident light. 

The n-region is patterned by electron-beam lithography into a 1~$\mu$m wide Hall bar along the [1$\bar{1}$0] crystal axis. The effective width of individual Hall contacts for local spin detection is 50-100~nm and separation between neighboring Hall crosses is  2~$\mu$m. Electrical gates controlling the spin-currents are placed between one or more pairs of the Hall crosses. The gates are realized by the p-type surface layer areas of the heterostructure which were locally masked and remained unetched during the fabrication of the n-channel Hall bar \cite{Kaestner:2007_a}. The laser beam is focused to a $\sim$1-2~$\mu$m spot at the lateral p-n junction or near the junction on the p-side of the epilayer.

In Fig.~1 we show experimental results measured at 4~K on a control device in which we did not pattern the gate electrodes. These measurements extend previous demonstration of the spin injection Hall effect in similar ungated structures \cite{Wunderlich:2008_a}. In the previous work we observed that injected spin-polarized electrical currents produce Hall effect signals which are proportional to the out-of-plane component of the local spin polarization. We also demonstrated that spins precess along the channel which results in spatially varying magnitude and sign of the Hall signals on several successive Hall crosses. Because of the limited number of discrete detection points these experiments did not provide a detailed picture of the spin precession of injected electrons. To better visualize the effect  we utilize here the optical activity of the device presented in Fig.~1 which extends over a several $\mu$m range from the lateral p-n junction into the unetched p-type side of the epilayer. By shifting the focused laser spot we can smoothly change the position of the spin injection point with respect to the detection Hall crosses. This  results in damped oscillatory Hall resistance,  $R_H=V_H/I_{PH}$, measured at each of the two successive Hall crosses labeled as H1 and H2 in Fig.~1. ($V_H$ is the Hall voltage and $I_{PH}$ is the photocurrent.) The oscillations at each Hall cross and the phase shift between signals at the two Hall crosses are consistent with a micron-scale spin precession period and with a spin-diffusion length which extends over more than one precession period.

Experiments in Fig.~1 are performed in two distinct electrical measurement configurations. In panel 1(a) we show data obtained with the source and drain electrodes at the far ends of the p and n-type sides of the lateral junction, respectively. In this geometry, spin-polarized electrical currents reach the detection Hall crosses and, therefore, we measure the spin-injection Hall effect (SiHE) signals \cite{Wunderlich:2008_a}. In panel 1(b) the electrical current is drained 2~$\mu$m before the first detection Hall cross H1. In this case only pure spin-current reaches  crosses H1 and H2 and we detect the inverse spin Hall effect (iSHE) signals \cite{Valenzuela:2006_a,Bruene:2008_a}.  The experiments in Fig.~1 demonstrate that in our 2DEG micro-channel we can realize Hall effect detection of injected spin-polarized electrical currents, as well as pure spin currents which do not produce Joule heating.

The conventional field-effect transistor functionality in our 2DEG channel achieved by the p-layer top gate is demonstrated in Fig.~2(a) where we show the gate voltage dependence of the channel current and mobility underneath the gate (see also the schematics of the measurement setup in panel 2(a)). At zero gate voltage we detect only a small residual channel current consistent with the large degree of depletion of the 2DEG in the unetched part of the epilayer. By applying forward or reverse voltages of an amplitude less than 1~V we can open or close the 2DEG channel, respectively, at negligible gate-channel leakage current. Within the range of measurable  signals we detect gate voltage induced changes of the channel current by 5 orders of magnitude while the mobility changes by 2 orders of magnitude. The main effect of the gate voltage  on the channel current is therefore via direct charge depletion/accumulation of the 2DEG but mobility changes are also significant. With increasing reverse gate voltage the mobility decreases because the 2DEG is shifted closer to the ionized donors on the other side of the AlGaAs/GaAs heterojunction and because screening of the donor impurity potential by the 2DEG decreases with depletion. 

The main result of our work, shown in Fig.~2(b), is the sensitivity of the measured Hall signal at the cross placed behind the gate on the voltage applied to the gate electrode. In order to exclude any potential gate voltage dependence of spin-injection conditions in our device the experiments are performed in the iSHE geometry with the electrical current drained before the gated part of the channel (see schematics in panel 2(b)). The data show two regimes of operation of our spin transistor. At large reverse voltages the Hall signals disappear as the diffusion of spin-polarized electrons from the injection region towards the detecting Hall cross is blocked by the repulsive potential of the intervening gate electrode. Upon opening the gate, the Hall signal first increases in analogy to the operation of the conventional field-effect transistor. We emphasize, however, that while the optically generated current $I_{PH}$ is kept constant, the electrical current in our experiments in the manipulation and detection parts of the transistor channel remains zero at all gate voltages. The onset of the output transverse electrical signal upon opening the gate is due to a pure spin current. The mechanism by which the spin-current generates the output  signal is also not due to a normal charge Hall effect because of the absence of magnetic field and charge current underneath the cross, but due to the iSHE. 

The initial increase of the detected output signal upon opening the gate is followed by a non-monotonic gate voltage dependence of the Hall voltage as shown in Fig.~2(b). This is in striking contrast to the monotonic increase of the normal electrical current in the channel observed in the conventional field-effect-transistor measurement in panel 2(a). Apart from blocking the spin-current at large reverse gate voltages, the intermediate gate electric fields  are modifying spin precession of the injected electrons and therefore the local spin polarization at the detecting Hall cross when the channel is open. This is the spin manipulation regime analogous to the original Datta-Das proposal of a spin transistor. The presence of this regime in our device is further demonstrated by comparing two measurements shown in Fig.~2(b), one where the laser spot is directly aligned at the lateral p-n junction (red, solid line) and the other one with the spot shifted by approximately 1~$\mu$m in the direction away from the detecting Hall crosses (black, dashed line). The reverse voltage at which the Hall signals disappear is the same in the two measurements because this effect is due to pinching off the channel underneath the gate. For gate voltages at which the channel is open, the signals are shifted with respect to each other in the two measurements, have opposite sign at certain gate voltages, and the overall magnitude of the signal is larger for smaller separation between injection and detection points, all confirming the spin precession origin of the observed effect. 

One of the important attributes of our non-destructive spin detection method integrated, together with the electrical spin manipulation, along the  semiconductor channel is the possibility to fabricate devices with a series of Hall cross detectors and also with a series of gates. In Fig.~3 we demonstrate the feasibility of this concept and of the ensuing logic functionality on a spin Hall effect transistor structure with two gates, first placed before cross H1 and second before H2. The scanning electron micrograph of the device is shown in panel 3(a). The measured data plotted in panel~3(b) demonstrate  that Hall cross H1 responds strongly to the electric field on the first gate, with a similar gate voltage characteristics as observed in the single-gate device in Fig.~2. As expected for the relative positions of the injection point, of Hall cross H1, and of the two gates in the device, the dependence of the signal at cross H1 on the second gate is much weaker. On the other hand, Hall cross H2 responds strongly to both gates, as shown in Fig.~3(c). Before the spin can reach the detecting Hall cross H2 it is manipulated by two external parameters. This is analogous to the measurement in Fig.~2(b) in which the position of the injection point played the role of the second parameter. The analogy between results in Figs.~2(b) and 3(c) provides further demonstration of the spin origin of the functionality of our transistor structures. 

In Fig.~3(d) we  demonstrate the AND logic function by operating both gates and by measuring the Hall electrical signal at cross H2. Intermediate gate voltages on both gates represent the input value ¬1¬ and give the largest electrical signal at H2 (positive for $\sigma^-$ helicity of the incident light), representing the output value 1. By applying to any of the two gates a large reverse (negative)  gate voltage, representing input ¬0¬, the electrical signal at H2 disappears, i.e., the output is ¬0¬. We again emphasize that this spin AND logic function is realized in the part of the semiconductor channel with zero electrical current.

We now proceed to the theoretical analysis of the measured data. First we characterize the transport regime in which our devices operate. The 2DEG mobilities in the etched, n-type part of the wafer and underneath the p-layer gates are $\lesssim3\times10^{3}$~cm$^2$/Vs, corresponding to mean-free-path $\lesssim10^{2}$~nm. This is much smaller than the precession length and the length of our 2DEG channel, i.e., the experiments are done in the diffusive, strong disorder weak spin-orbit coupling regime. As explained in Ref.~\cite{Wunderlich:2008_a}, the Hall effect and the spin-precession effect can be decoupled in this regime. The  Hall effect measures  the local out-of-plane component of the spin polarization of carriers and originates from the spin-orbit coupling induced skew scattering \cite{Wunderlich:2008_a}.  In the following analysis we focus on the spin-precession and spin-diffusion lengths. The possibility to observe and utilize spin precession of an ensemble of electrons in the diffusive regime is demonstrated by our numerical Monte Carlo simulations \cite{Wunderlich:2008_a,Liviu} shown in Fig.~1(c).

The numerically obtained spin-precession period is well described by an analytical formula derived from the dynamics of the spin-density matrix \cite{Liviu}, $L_{SO}=\pi\hbar^2/m^*(|\alpha|+|\beta|)$.
There are two regimes in which spin precession can be observed in the diffusive transport regime.
In one regime the width of the channel is not relevant and spin-diffusion length larger than the precession length  occurs due to the single-particle transport analogue of the spin Helix state \cite{Bernevig:2006_a} realized at 2DEG Rashba and Dresselhaus spin-orbit fields of equal or similar strengths, $\alpha\approx-\beta$ for our bar orientation. ($m^*=0.067$ in the above equation is the electron effective mass in GaAs.) When the two spin-orbit fields are not tuned to similar strengths, the spin-diffusion length is approximately given by $\sim L_{SO}^2/w$  and spin-precession is therefore observable only when the width $w$ of the channel is comparable or smaller then the precession length \cite{Liviu,Kiselev:2000_a,Kettemann:2007_a}. 

Since our semiconductor heterostructure has a complex design to provide simultaneously the means for spin injection, electrical gating, and detection we did not rely on further fine tuning of the internal spin-orbit fields to realize the spin Helix state condition. Instead we fabricated narrow Hall bars whose width is smaller than the precession length and used a strongly focused light spot for spin-injection. As shown in Fig.~1(c), several precessions are readily observable in this quasi 1D geometry even in the diffusive regime and for $\alpha\neq-\beta$ ($\alpha=5.5$~meV\AA, $\beta=-24$~meV\AA\, in the simulations in Fig.~1(c)). We also demonstrate in Fig.~1(c) that the spin-precession and spin-diffusion lengths are independent in this regime of the mean-free-path, i.e., of the mobility of the 2DEG channel \cite{Liviu}.

The strength of the confining electric field of the 2DEG underneath the gate changes by up to a factor of $\sim$2 in the range of applied gate voltages in our experiments. It implies \cite{Wunderlich:2008_a} comparably large changes in the strength of the internal spin-orbit fields in the 2DEG channel. The dependence on the spin-orbit field strength shown in the above equation and confirmed by MC simulations \cite{Liviu} (and the independence on the momentum of injected electrons) implies also comparably large changes of the spin-precession length. These estimates corroborate the observed spin-manipulation in our spin Hall effect transistors by external electric fields applied to the gates. We also point out that while spins can be manipulated  by low electric gate voltages $\sim 1$~V, the Hanle effect spin-precession due to external magnetic field is not observable in 2DEGs with typical strength of the spin-orbit coupling \cite{Wunderlich:2008_a} even at magnetic fields as large as $\sim 1$~T.

To conclude, we have demonstrated the semiconductor spin transistor  in which spin manipulation and detection 
are both realized electrically via spin-orbit coupling phenomena in a part of the transistor channel with zero electrical current. We have shown that logic operation can be performed in a diffusive spin Hall channel with multiple gates. The approach to spin transistors introduced in this paper has no fundamental physics limitations on high temperature operation of the devices.

\begin{figure}[h]
\vspace*{-3cm}
\hspace*{-0cm}\includegraphics[scale=0.6]{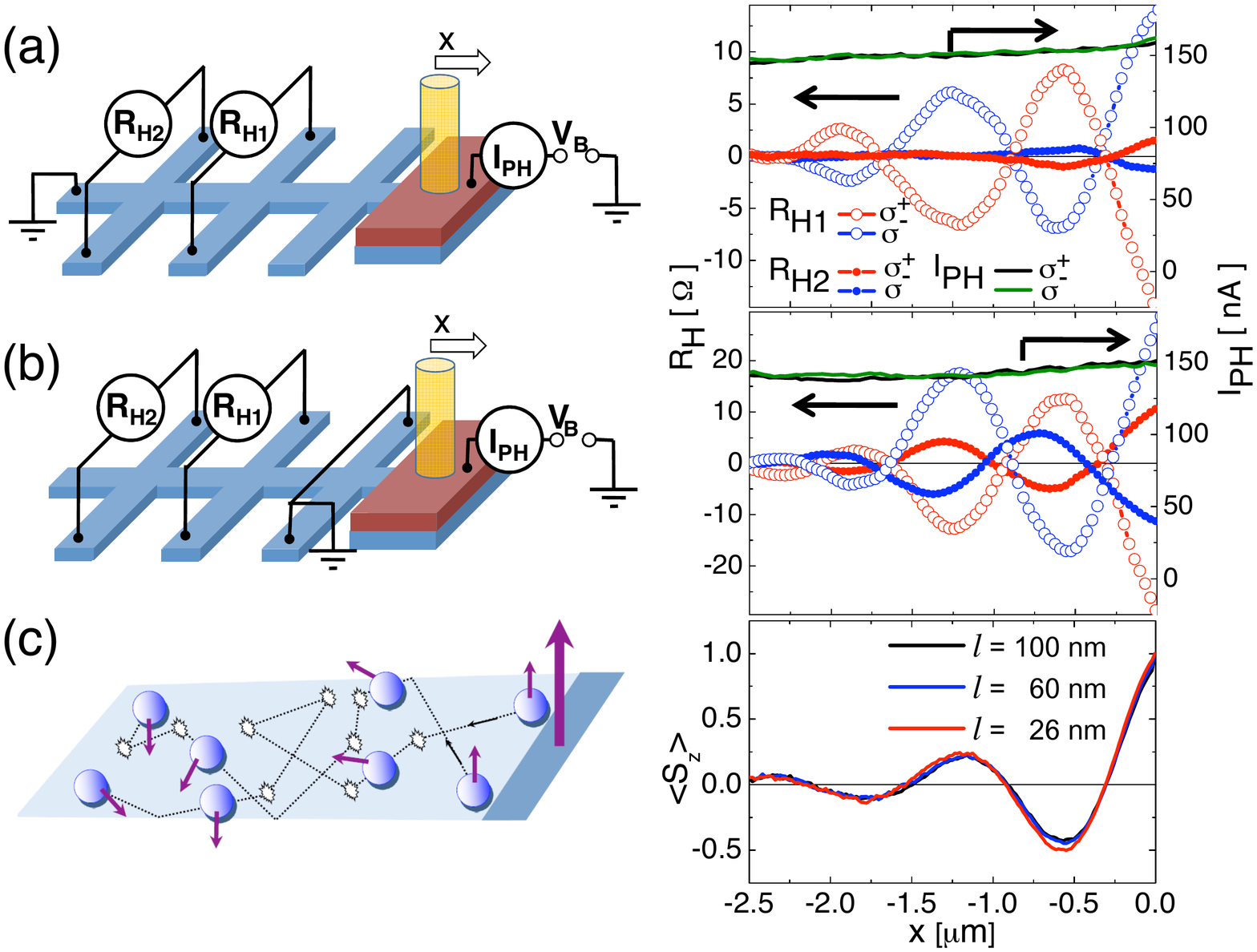}
%

\caption{(a) Schematics of the measurement setup with optically injected spin-polarized electrical current propagating through the Hall bar and corresponding experimental Hall effect signals at crosses H1 and H2. The Hall resistances, $R_H=V_H/I_{PH}$, for the two opposite helicities of the incident light are plotted as a function of the focused ($\sim$1~$\mu$m) light spot position, i.e., of the position of the injection point. The $x$-axis represents the relative shift of the spot and increasing amplitude of $x$ corresponds to shifting the spot further away from the Hall detectors. (The focused laser beam is indicated by the yellow cylinder in the schematics.) The transverse electrical signals at each cross change sign with changing helicity of the light, i.e., with changing spin-polarization of the injected electrons, consistent with the SiHE. Opposite sign of the signal at the two successive Hall crosses detected at a given position of the light spot demonstrate the spin precession along the semiconducting channel. The precession is also demonstrated by the oscillatory pattern measured at each cross as a function of the position of the focused laser spot. As also shown in the plot, the optical current $I_{PH}$ is independent of the helicity of the incident light and varies only weakly with the light spot position. (b) Same as (a) for measurement geometry in which electrical current is closed before the first detecting Hall cross H1, i.e., the measured transverse signals at cross H1 and H2 correspond to the iSHE. (c) Monte-Carlo simulations of the out-of-plane component of the spin of injected electrons averaged over the 1~$\mu$m bar cross-section assuming Rashba field $\alpha=5.5$~meV\AA, Dresselhaus field $\beta=-24$~meV\AA, and different values of the mean-free-path $l$.
}
\label{f1}
\end{figure}
\begin{figure}[h]
\hspace*{-0cm}\includegraphics[scale=0.6]{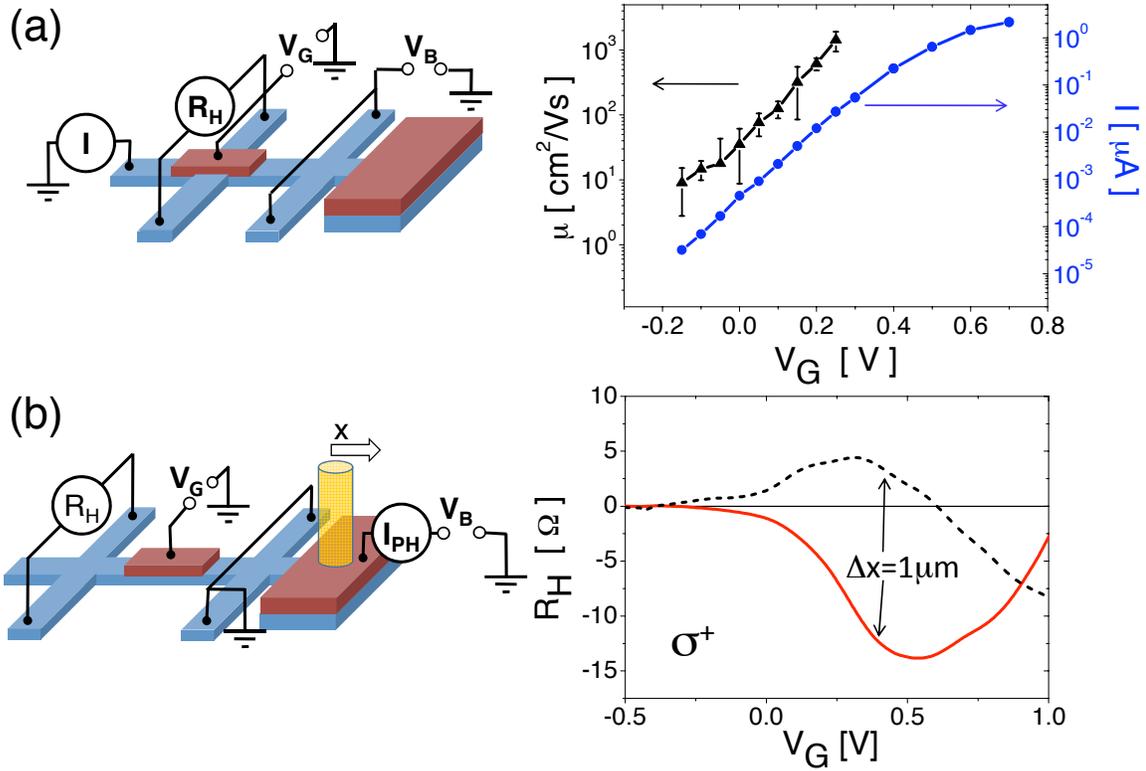}
%

\caption{(a) Schematics of the measurement setup corresponding to the conventional field-effect transistor and experimental dependence of the electrical current (blue) through the channel and mobility (black) underneath the gate on the gate voltage. (b) Schematics of the setup and iSHE measurements as a function of the gate voltage at a Hall cross placed behind the gate electrode for two light spot positions with a relative shift of 1~$\mu$m and the dashed black curve corresponding to the spot shifted further away from the detection Hall cross. The data demonstrate the realization of the spin Hall effect transistor. 
}
\label{f2}
\end{figure}

\begin{figure}[h]

\hspace*{-1cm}\includegraphics[scale=0.65]{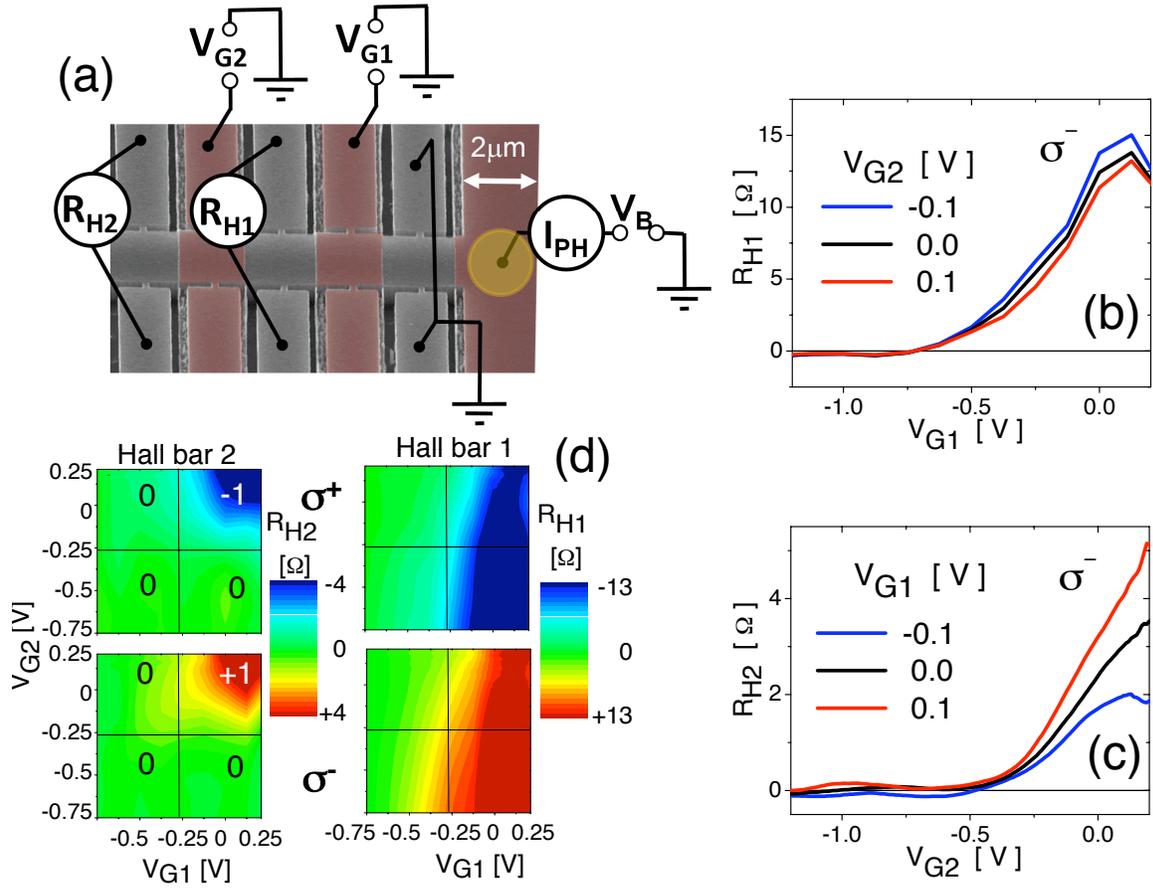}
%

\caption{(a) Scanning electron micrograph image and schematics of the device with two detecting Hall crosses H1 and H2 and one gate placed before cross H1 and the second gate placed behind cross H1 and before cross H2. Gates and p-side of the lateral p-n junction are highlighted in red. The focused laser beam is indicated by the yellow spot. (b) Hall signals at cross H1 measured as a function of the first gate voltage. These gating characteristics are similar to the single-gate device in Fig.~2(b) and have much weaker dependence on the second gate voltage.  (c) Hall signals at cross H2 measured as a function of the second gate voltage. The curves show strong dependence on the voltages on both gates. (d) Demonstration of the spin AND logic function by operating both gates (input signals) and measuring the response at Hall cross H2 (output signal). Measured data at cross H1 are also shown for completeness. 
}
\label{f3}
\end{figure}

\end{document}